%% file: paper_forArxiv.tex
\documentclass[conference]{IEEEtran}
\IEEEoverridecommandlockouts

\usepackage{cite}
\usepackage{amsmath,amssymb,amsfonts}
\usepackage{algorithmic}
\usepackage{graphicx}
\usepackage{textcomp}
\usepackage{xcolor}
\usepackage{array}
\usepackage{hyperref}
\usepackage{cleveref}
\usepackage{longtable}
\usepackage{balance}

\usepackage[many]{tcolorbox}

\newtcolorbox{fancybox}[1][]{
  enhanced,
  attach boxed title to top center={yshift=-3mm,yshifttext=-1mm},
  colback=blue!15!white,
  colframe=blue!25!black,
  colbacktitle=blue!25!black,
  fonttitle=\bfseries,
  title=#1,
  boxed title style={size=small, colframe=blue!65!black, colback=blue!65!white},
  drop fuzzy shadow,
  width=0.5\textwidth,
  breakable
}

\newtcolorbox{RQsTakeAwaybox}[1][]{
  enhanced,
  attach boxed title to top center={yshift=-3mm,yshifttext=-1mm},
  colback=green!25!white,
  colframe=green!45!black,
  colbacktitle=green!45!black,
  fonttitle=\bfseries,
  title=#1,
  boxed title style={size=small, colframe=green!45!black, colback=green!35!black},
  drop fuzzy shadow,
  width=0.49\textwidth,
  breakable
}

\usepackage{listings}
\newcommand{\sectopic}[1]{\vspace{0.2em}\par\noindent{\textit{\bfseries #1}}}

\begin{document}

\title{Optimizing Large Language Model Hyperparameters for Code Generation}


\author{\IEEEauthorblockN{Chetan Arora}
\IEEEauthorblockA{
\textit{Monash University}\\
Melbourne, Australia \\
chetan.arora@monash.edu}
\and
\IEEEauthorblockN{Ahnaf Ibn Sayeed}
\IEEEauthorblockA{
\textit{Monash University}\\
Melbourne, Australia \\
ahnafibnsayeed@gmail.com}
\and
\IEEEauthorblockN{Sherlock Licorish}
\IEEEauthorblockA{
\textit{University of Otago}\\
Dunedin, New Zealand \\
sherlock.licorish@otago.ac.nz}
\and
\IEEEauthorblockN{Fanyu Wang}
\IEEEauthorblockA{
\textit{Monash University}\\
Melbourne, Australia \\
fanyu.wang@monash.edu}
\and
\IEEEauthorblockN{Christoph Treude}
\IEEEauthorblockA{
\textit{Singapore Management University}\\
 Singapore \\
ctreude@smu.edu.sg}
}

\maketitle
\thispagestyle{plain}
\pagestyle{plain}
\pagenumbering{arabic}

\begin{abstract}
Large Language Models (LLMs), such as GPT models, are increasingly used in software engineering for various tasks, such as code generation, requirements management, and debugging. While automating these tasks has garnered significant attention, a systematic study on the impact of varying hyperparameters on code generation outcomes remains unexplored. This study aims to assess LLMs' code generation performance by exhaustively exploring the impact of various hyperparameters. Hyperparameters for LLMs are adjustable settings that affect the model's behaviour and performance. Specifically, we investigated how changes to the hyperparameters—temperature, top probability (top\_p), frequency penalty, and presence penalty—affect code generation outcomes. We systematically adjusted all hyperparameters together, exploring every possible combination by making small increments to each hyperparameter at a time. This exhaustive approach was applied to 13 Python code generation tasks, yielding one of four outcomes for each hyperparameter combination: no output from the LLM, non-executable code, code that fails unit tests, or correct and functional code. We analysed these outcomes for a total of 14,742 generated Python code segments, focusing on correctness, to determine how the hyperparameters influence the LLM to arrive at each outcome. Using correlation coefficient and regression tree analyses, we ascertained which hyperparameters influence which aspect of the LLM. Our results indicate that optimal performance is achieved with a temperature below $0.5$, top probability below $0.75$, frequency penalty above $-1$ and below 1.5, and presence penalty above $-1$. We make our dataset and results available to facilitate replication.
\end{abstract}


\input{Files/introduction}
\input{Files/empirical_settings}
\input{Files/results}

\input{Files/discussion}

\input{Files/threats}

\input{Files/related}
\input{Files/conclusion}

\balance
\bibliographystyle{IEEEtran}
\bibliography{refs.bib}

\end{document}

%% file: Files/introduction.tex
\section{Introduction} \label{sec:introduction}
Large Language Models (LLMs), such as GPT models, have revolutionized the field of software engineering (SE) by automating various tasks, including code generation~\cite{hou2023large}, software testing~\cite{nguyen2023generative}, requirements and process management~\cite{arora2024advancing}. The ability of these models to assist in both complex and routine SE tasks has made them invaluable tools for software developers, promising significant gains in productivity and efficiency~\cite{hou2023large,nguyen2023generative,al2024enhancing}. 

Given that code generation is one of the most prevalent use cases of LLMs in software engineering, there have been several efforts to enhance the performance of LLMs for code generation, both in the research domain~\cite{jiang2023selfplanning,li2023think} and by bigger corporations, e.g., the advent of various coding assistants such as Microsoft 365 Copilot\footnote{\url{https://www.microsoft.com/en-au/microsoft-365/microsoft-copilot}} and GitHub Copilot\footnote{\url{https://github.com/features/copilot}}. The research efforts include, developing custom frameworks for complex code generation~\cite{li2023think} and fine-tuning moulded or incorporating external tools to boost the performance of LLMs in code generation~\cite{chen2021evaluating}.

Despite these advancements, one critical aspect of LLMs that remains under-explored is the systematic study of hyperparameters and their impact on code generation outcomes. Hyperparameters, which are adjustable settings that affect the LLMs' behaviour and performance, can significantly influence the quality of the generated text (or code). Precise fine-tuning of these hyperparameters, more often than not, leads to better results. For example, in ML algorithms, hyperparameter optimization is considered one of the key steps for training a model and is part of the best practices~\cite{feurer2019hyperparameter}. Hyperparameter optimization of LLMs has also led to better results in other domains, e.g., text summarization~\cite{alexandr2021fine} and cost management of text generation~\cite{wang2023costeffectivehyperparameteroptimizationlarge}. However, no study (to the best of our knowledge) has studied the impact of varying hyperparameters for code generation (or SE in general), and exploring the best configuration that leads to optimum results. Existing SE research has predominantly focused on model architectures and training methodologies, often overlooking the tuning of hyperparameters or studying the impact of hyperparameter settings, if any, on the code generation task.

To address this gap, our study aims to systematically assess the performance of LLMs in code generation by exhaustively exploring the impact of several key hyperparameters~\cite{wang2023costeffectivehyperparameteroptimizationlarge}. Specifically, we focus on temperature, top probability (top\_p), frequency penalty, and presence penalty settings\footnote{\url{https://platform.openai.com/docs/api-reference}}.
Temperature controls the randomness of the LLM's output, influencing the diversity of the generated text (or code, in our case). Top probability (top\_p) limits the sampling pool of tokens, affecting the creativity and variability of the responses. Frequency penalty reduces the likelihood of repeating tokens, encouraging more novel outputs, while the presence penalty discourages the model from generating tokens that have already appeared, promoting diversity in the generated code. 
By methodically adjusting these hyperparameters and evaluating their effects across multiple code generation tasks, we aim to identify configurations that yield the best outcomes and those that should be avoided to prevent suboptimal performance. We note that there are some other hyperparameters, e.g., $max\_tokens$, which fixes the maximum number of tokens generated, and $logit\_bias$ that allows to adjust the likelihood of specific tokens being selected or suppressed during text generation, effectively influencing the model's output. We did not experiment with these hyperparameters to avoid constraining the LLMs in terms of the length of the response generated or artificially influencing the exact phrases in the output.

\sectopic{Contributions.} This study makes several key contributions to the field of SE and the use of Large Language Models (LLMs) for code generation. First, guided by the following two key research questions (RQs), this is the first study (to the best of our knowledge) to systematically investigate the impact of hyperparameters on code generation via LLMs.

\textbf{RQ1. Does variation in LLM hyperparameters lead to an impact on the quality of code generated by LLMs?}

\textbf{RQ2. What are the optimal hyperparameter settings for achieving the best code generation outcomes with LLMs?}


To address these questions, we conducted an exhaustive evaluation of hyperparameter combinations of GPT-3.5 Turbo model~\footnote{\url{https://platform.openai.com/docs/models}} on 13 Python code generation tasks. Each combination was tested to observe one of four possible outcomes: no output, output generated but non-executable code, code that fails basic unit tests, or correct and functional code. Through exhaustive experimentation and analysis, we identify specific hyperparameter configurations that consistently yield the best (or suboptimal) code generation outcomes. These settings can serve as practical guidelines for developers and researchers seeking to optimize LLM performance for code generation tasks. We also make our results and dataset publicly available to facilitate replication\footnote{\url{https://figshare.com/s/3e179563580a629fa9c2}}.

\sectopic{Structure.}
Section~\ref{sec:empiricalEvaluationSettings} presents the empirical evaluation settings for our study. Section~\ref{sec:Results} presents the results of our RQs. Section~\ref{sec:discussion} provides a detailed discussion on our results and their implications. Section~\ref{sec:threats} reviews the threats to validity and limitations of our study. Section~\ref{sec:RelatedWork} positions our work against the related work and Section~\ref{sec:conclusion} concludes the paper.

%% file: Files/empirical_settings.tex
\section{Empirical Evaluation Settings} \label{sec:empiricalEvaluationSettings}
\subsection{Study Details} \label{methodology}
A study comparing the GPT model's code generation efficacy on \emph{ten} programming languages concluded that the model performed better on high-level dynamically typed languages such as Python~\cite{buscemi2023comparative}. Therefore, we decided to conduct our hyperparameter investigation with a dataset containing programming tasks in Python. We used an existing dataset by Feng et al.~\cite{10196869} as a basis for our evaluation. This dataset contains 72 Python programming tasks to evaluate the performance of ChatGPT on crowd-sourced social data. Our research study aims to add to this evaluation by using the programming tasks from the dataset to systematically investigate the impact of tuning various hyperparameters. 

To keep our project scope narrow and focused, we decided to evaluate the tasks on only one metric: functionality (i.e. correctness of the code), as it was the highest-rated metric for code evaluation by software developers \cite{NDUKWE2023111524}. Functionality tests for both syntactic and semantic correctness need to be correct for the code to be functional. We decided to test functionality using Python's `\textit{unittest}' library\footnote{https://docs.python.org/3/library/unittest.html}. The generated codes needed to be in a function format to use this library so that the methods in the `\textit{TestCase}' class could be called. To automate the LLM generation process, we decided to use OpenAI's API\footnote{https://platform.openai.com/docs/api-reference/}, and based all our experiments on \textsf{gpt-3.5-turbo} model. 

Due to our focus being code functionality and the need for the output to be functions (to utilise the '\textit{unittest}' library), we decided to filter out tasks from the dataset \cite{10196869} that could not be evaluated based on our criteria. For example, one of the ineligible tasks asked for pseudocode for an adaptive BCI algorithm. Another task was a question about how to avoid repetitions in a `for loop'. As a result, we narrowed the dataset down to 13 tasks (Table~\ref{tbl:Programs}). These 13 tasks were used to generate code in Python in a function format so that they could be tested for functionality using the `\textit{unittest}' library.

\begin{table}[]
\caption{Python Tasks Description}
\label{tbl:Programs}
\begin{tabular}{|l|m{7.5cm}|}
\hline
\textbf{ID} & \textbf{Description}\\ \hline \vspace{0.1em}
P1 & \vspace{0.1em}  \textbf{ApproximateSine:} This function receives a number as an input and outputs the approximate of the sine value of this number. \vspace{0.1em} \vspace{0.1em} \\ \hline \vspace{0.1em}
P2  & \vspace{0.1em} \textbf{ReverseString:} This function receives a list of strings, reverses each string and returns a new list with the reversed strings. \vspace{0.1em} \\ \hline \vspace{0.1em}
P3  & \vspace{0.1em}  \textbf{PrimeNumbers:} This function receives a number as input and outputs a list containing prime numbers less than or equal to the input. \vspace{0.1em} \\ \hline \vspace{0.1em}
P4  & \vspace{0.1em}  \textbf{FourthLargest:} This function receives an array as input and outputs the fourth largest element in the array. \vspace{0.1em} \\ \hline \vspace{0.1em}
P5  & \vspace{0.1em} \textbf{LargestSumListInFile:} This function reads the integers from a text file, sums the groups of integers, finds the group with the largest sum of values, and then outputs the group that had the largest sum as a list. In the text file, there are rows of numbers, and the groups are separated by a blank row. Therefore, all the numbers that are next to each other are in the same group, and the groups are differentiated by the blank rows. \vspace{0.1em} \\ \hline \vspace{0.1em}
P6 & \vspace{0.1em}  \textbf{FizzBuzz:} This function takes in an integer and outputs either Fizz, Buzz or FizzBuzz. \vspace{0.1em} \\ \hline \vspace{0.1em}
P7 & \vspace{0.1em}  \textbf{CompareIntegers:} This function compares to inputted integers and returns true if the first integer is larger. \vspace{0.1em} \\ \hline \vspace{0.1em}
P8 & \vspace{0.1em}  \textbf{WordCount:} This function counts the number of words in the input string and outputs the number of words. \vspace{0.1em} \\ \hline \vspace{0.1em}
P9  & \vspace{0.1em}  \textbf{TargetSum:} This function takes in a target number and a list of numbers as input and outputs a list of two indices from the list that sum to the target number. \vspace{0.1em} \\ \hline \vspace{0.1em}
P10 & \vspace{0.1em}  \textbf{EigenValue:} This function takes a list of matrices as input, calculates its determinant plus largest eigenvalue, and outputs the index of the matrix with the smallest of this value. \vspace{0.1em} \\ \hline \vspace{0.1em}
P11 & \vspace{0.1em}  \textbf{Fibonacci:} This function takes an integer as input and outputs a list of numbers from the fibonacci sequence. \vspace{0.1em} \\ \hline \vspace{0.1em}
P12 & \vspace{0.1em}  \textbf{InitialsTransform:} This function takes in a list of names as input and outputs a new list with only the initials from the names. \vspace{0.1em} \\ \hline \vspace{0.1em}
P13 & \vspace{0.1em}  \textbf{DoubleTheNumber:} This function takes an integer as input and outputs the integer multiplied by 2. \vspace{0.1em} \\ \hline
\end{tabular}
\end{table}

\subsection{Hyperparameter Tuning}\label{subsec:HyperparameterTuning}
The OpenAI API used in our study allows to tune several hyperparameters for generating outputs. These include,  temperature ($temp$), 
top probability ($top\_p$), frequency penalty ($freq\_pen$), presence penalty ($pres\_pen$), maximum tokens ($max\_tokens$) and logit bias ($logit\_bias$). As mentioned in Section~\ref{sec:introduction}, for the purpose of our research, we only explored hyperparameters that would be relevant to our RQs and would not artificially bias or limit the output of the LLMs. Therefore, hyperparameters like \textit{logit\_bias} and \textit{max\_tokens} were not changed. The four tuneable hyperparameters that are relevant to our RQs are:

\sectopic{Temperature ($temp$)} controls the randomness of the model's predictions. A lower $temp$ value makes the model's responses more deterministic and focused, while a higher $temp$ increases randomness and creativity~\cite{ouyang2023llm}. $temp$ ranges between [0.0, 2.0] in OpenAI API, \textsf{gpt-3.5-turbo} model.
For example, for a prompt \textit{``How do you create a list in Python?''}, for lower $temp$ values, the LLM would more consistently respond, whereas for higher $temp$ values, it will be more creative.
\begin{fancybox}[Lower $temp$]
\textit{To create a list in Python, you can use square brackets. For example: my\_list = [1, 2, 3].}
\end{fancybox}

\begin{fancybox}[Higher $temp$ ]
\textit{In Python, you can create a list by enclosing elements in square brackets, like: my\_list = [1, `apple', 3.14].}
\end{fancybox}

\sectopic{Top Probability, also known as Nucleus Sampling ($top\_p$)} determines the cumulative probability distribution for the next token \cite{ouyang2023llm}. $top\_p$ ranges between [0.0, 1.0] in OpenAI API, \textsf{gpt-3.5-turbo} model. Setting a value for top\_p restricts the model's choice to the most probable tokens whose cumulative probability exceeds the threshold ($p$). Setting top\_p to 0.5 means only the tokens comprising the top 50\% probability mass are considered. For example, if the LLM is generating the next word in the sentence, \textit{``The cat sat on the...''}. The model assigns probabilities to each potential next word:

\begin{fancybox}[Setting $top\_p$]
\textbf{Probabilities}:  mat    (0.4), sofa   (0.3), floor  (0.2), table  (0.05), roof   (0.03), carpet (0.02).

If top\_p is set to 0.9, the model selects words from the smallest subset of possible words whose cumulative probability exceeds 0.9. The model will only consider ``mat'', ``sofa'' and ``floor'', and exclude ``table'', ``roof'' and ``carpet'' because including them would exceed the cumulative probability of 0.9.
\end{fancybox}
    
\sectopic{Frequency Penalty ($freq\_pen$)} penalises the model for generating frequently used tokens. When this hyperparameter is set to high, the model is encouraged to generate less common or more diverse output or reduces the likelihood of repeating tokens that have already appeared in the response, encouraging more novel outputs~\cite{bhavya2022analogy}. $freq\_pen$ ranges between [-2.0, 2.0] in OpenAI API, \textsf{gpt-3.5-turbo} model. For example, if we are generating text for , \textit{``Python is a popular programming language because...''}. The frequency penalty can influence how often certain words are repeated in the generated text. With a high frequency penalty, the model avoids repeating the word ``popular'' and instead uses synonyms or rephrases the sentences to convey the same meaning without redundancy.

\begin{fancybox}[Lower $freq\_pen$]
\textit{Python is a popular programming language because it is easy to learn. It is popular because it has a large community. Python is popular because it is versatile.}
\end{fancybox}

\begin{fancybox}[Higher $freq\_pen$]
\textit{Python is a popular programming language because it is easy to learn. It has a large community and is known for its versatility.}
\end{fancybox}

\sectopic{Presence Penalty ($pres\_pen$)} reduces the likelihood of a token being generated if it has already appeared at all in the generated text. $pres\_pen$ ranges between [-2.0, 2.0] in OpenAI API, \textsf{gpt-3.5-turbo} model. With a higher presence penalty, the model is discouraged from repeating the same words or phrases~\cite{bhavya2022analogy}.

\begin{fancybox}[Zero $pres\_pen$]
\textit{Python is a popular programming language because it is popular. Python is used widely and is popular because it is easy to learn. Python's popularity is also due to its large community.}
\end{fancybox}

\begin{fancybox}[Higher $pres\_pen$]
\textit{Python is a popular programming language because it is widely recognised. It is preferred for its ease of use and the strong support from its community. Python also offers great versatility and numerous applications.}
\end{fancybox}

In the API documentation\footnote{https://platform.openai.com/docs/api-reference/}, OpenAI has recommended altering either \textit{temperature} or \textit{top\_p}, but not both \cite{döderlein2023piloting}. Therefore, we performed two sets of experiments to explore the different hyperparametric configurations. The first set iterated through $temp$, $freq\_pen$ and $pres\_pen$, and the second set iterated through $top\_p$, $freq\_pen$ and $pres\_pen$.
We decided to test each task systematically with all combinations of hyperparameter configurations. We decided to use increments of 0.25 for $temp$ and $top\_p$. We posit that the increments of 0.25 provide a sufficient level of detail to observe meaningful changes in model performance. This allows us to identify trends and optimal settings without overlooking significant variations. Finer increments, such as 0.1, would exponentially increase the number of combinations to investigate, leading to significantly higher computational costs and longer experiment times. We also note that our initial experiments at increments of 0.1 did not lead to significantly different results. Hence, by using 0.25 increments, we keep the computational load manageable while still obtaining a comprehensive understanding of the hyperparameter space. Along similar lines, we used increments of 0.5 for $freq\_pen$ and $pres\_pen$. Hence, we had \emph{nine} combinations for $temp$ = [0.0, 0.25, 0.5, 0.75, 1.0, 1.25, 1.5, 1.75, 2.0], five combinations for $top\_p$ = [0.0, 0.25, 0.5, 0.75, 1.0], and \emph{nine} combinations each for $freq\_pen$ and $pres\_pen$ = [-2.0, -1.50, -1.0, -0.50, 0.0, 0.50, 1.0, 1.50, 2.0]. As mentioned above, we did not vary $temp$ and $top\_p$, based on OpenAI recommendations. Hence, we experimented with 9*9*9 = 729 combinations (hereafter, referred to as $SET_{temp}$), when varying $temp$, $freq\_pen$ and $pres\_pen$, and 5*9*9 = 405 combinations (hereafter, referred to as $SET_{topp}$), when varying $top\_p$, $freq\_pen$ and $pres\_pen$. In total, we experimented with 1,134 combinations of hyperparameter settings. We applied each of these settings for 13 tasks, hence, in total, experimenting on the model with 14,742 combinations.

\subsection{Analysis Procedure}~\label{subsec:Analysis}
Given the large number of combinations, we had to automate the process of analyzing the generated code and its validity. We first generated each task code inside a function in order to be unit tested using the Python \textit{unittest} library. We further note that the original prompts from the baseline study~\cite{10196869} did not instruct the model to output a function that would take inputs and return outputs. Hence, we tweaked the original prompts slightly to account for this. Further, we had to automatically test each function, and thus, we manually wrote some unit test cases for each of these tasks. These were written by the second author and validated by two co-authors, who have several years of software engineering and software testing experience, including more than five years of industry experience. Another factor to consider was how the model would name the functions in its outputs. It was observed that without explicitly mentioning a function name in the prompts, the resulting functions would often have different names from each other. For instance, the prompt \textit{``Can you provide a sample Python code to approximate the sine function?''} resulted in varying functions with names \textit{sin\_approx}, \textit{approximate\_sine} and \textit{approximateSine}. This could prove a challenge when unit testing, as the function names used in the unit test step, would not match the names in all the outputs. Therefore, the prompts had to be modified to ask for a function and specify the function's name so that all the outputs from the LLM have the same function name.

When modifying the original prompts in the dataset, we wanted to limit our changes and keep the prompts as close to the original as possible. We also wanted to avoid biasing the LLM to potentially alter its outputs. The highest chances of this happening would be when we would give it specific function names. Therefore, we wanted to be cautious and did not want to use function names that could potentially change how the LLM behaves. In an effort to avoid biasing the LLM, we decided to use the '\textit{camelCase}' naming convention, as research has shown it to be more readable \cite{5090039}. For the prompt described in the previous paragraph, we changed it to the following prompt: \textit{``Can you provide a sample python function 'approximateSine' to approximate the sine function?''}. We performed some initial testing with this prompt and compared the outputs with the unedited prompt, and the logic of the generated solutions was identical for both prompting styles.

For some of the hyperparameter configurations, when sending requests to OpenAI, the model would not output anything and instead raise an error. This error would always contain the following message: \textit{``Failed to create completion as the model generated invalid Unicode output. Unfortunately, this can happen in rare situations. Consider reviewing your prompt or reducing the temperature of your request.''} As a result, we decided to include this as one of the outcomes in our study.

For labelling the results of our experiments, we observed that each experiment would achieve one of the following four outcomes.
\begin{itemize}
    \item $Code_0$: the generated solution is functionally correct and passes all unit tests.
    \item $Code_1$: The model fails to generate a solution.
    \item $Code_2$: The generated solution is not executable.
    \item $Code_3$: The generated solution fails (one+) unit tests.
\end{itemize}


%% file: Files/results.tex
\section{Results} \label{sec:Results}

\begin{figure}[htbp]
    \centering
    \includegraphics[width=\columnwidth]{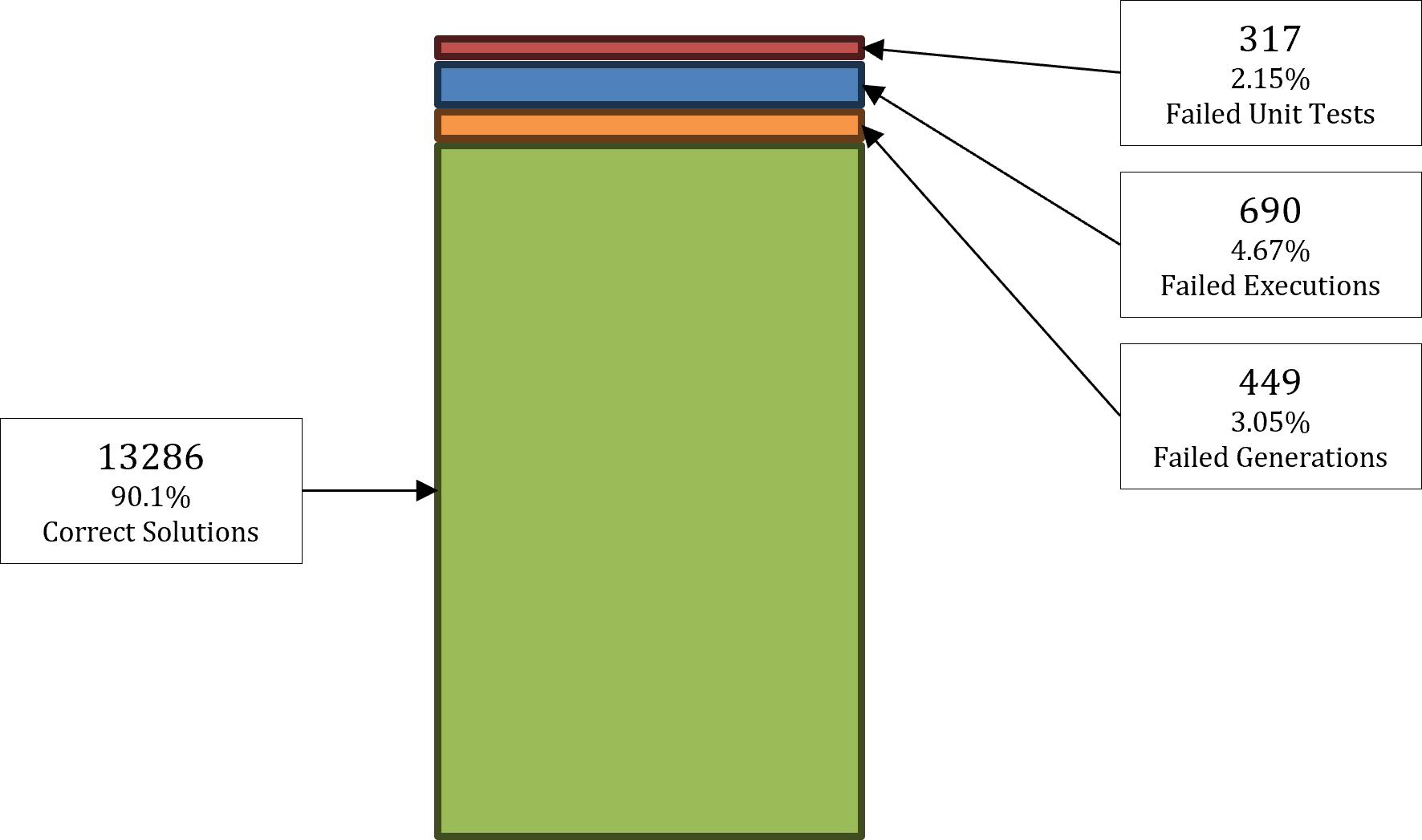}
    \caption{Experiment Outcomes Bar Graph}
    \label{fig:bargraph}
\end{figure}

In this section, we present the overall results of our experiments for our RQs. Fig.~\ref{fig:bargraph} shows the overall aggregated results of the four outcomes of all our 14,742 combinations of code generated. Overall, 90.1\% of the code snippets generated were correct ($Code_0$), which shows the overall efficacy of LLMs in generating solutions for simple coding tasks. The incorrect solutions constituted 9.9\%, i.e., with 3.05\% failing to generate ($Code_1$), 4.67\% failing to execute ($Code_2$), and 2.15\% failing unit tests ($Code_3$). These results show that in $\approx$10\% cases, the hyperparameter tuning might make a difference.  

\subsection{RQ1 Analysis} \label{subsec:RQ1Analysis}
Next, to answer RQ1 and obtain a high-level perception of how each hyperparameter affects each outcome type, we calculated 16 correlation coefficients ($4*4=16$). Since we have two sets of data ($SET_{temp}$ where the temperature is configured and $top\_p$ is left at default, and $SET_{topp}$ where $top\_p$ is configured and $temp$ is left at default) and both sets configure $freq\_pen$ and $pres\_pen$. We computed the weighted average correlation coefficients for $freq\_pen$ and $pres\_pen$. For $temp$ and $top\_p$ however, we could use the coefficients from their $SET_{temp}$ and $SET_{topp}$, respectively. The resulting correlation coefficients are presented in \autoref{tab:corrcoef}, computed using Pearson correlation coefficient~\cite{cohen2009pearson}. A larger coefficient value (range between -1 and +1) indicates a stronger correlation between the input and output variables. A positive coefficient suggests that increasing the input variable leads to an increase in the output variable, whereas a negative coefficient indicates that increasing the input variable results in a decrease in the output variable. The hyperparameters that moderately (0.3 - 0.5, or 0.5 to -0.3) or significantly (0.5 - 1.0 or  -1 to -0.5) impacted the outcomes have their correlation coefficients in bold in \autoref{tab:corrcoef}.

\begin{table}[!htb]
\caption{Correlation Coefficients of each hyperparameter against each outcome type}
\label{tab:corrcoef}
\centering
\begin{tabular}{|c|c|c|c|c|}
\hline
Outcome Types      & $Code_0$ & $Code_1$ & $Code_2$ & $Code_3$ \\ \hline
Temperature        & \textbf{-0.721}  & \textbf{0.549}   & \textbf{0.408}   & \textbf{0.418}  \\ \hline
Top Probability    & \textbf{-0.361}  & N/A       & 0.169   & \textbf{0.426}  \\ \hline
Frequency Penalty  & 0.058   & 0.188  & -0.257  & -0.046 \\ \hline
Presence Penalty   & 0.030   & 0.073  & -0.028  & -0.028 \\ \hline
\end{tabular}
\end{table}

$temp$ had the most significant effect on all four outcomes, particularly on $Code_0$ (code correctness), with a correlation coefficient of -0.721. This strong negative correlation suggests that lowering the temperature leads to an increase in code correctness, thereby improving overall code generation quality. $top\_p$ also demonstrated a notable impact, with a correlation coefficient of -0.361, indicating an inverse relationship with correctness similar to that of temperature. In contrast, $freq\_pen$ and $pres\_pen$ had minimal effect on the outcome.

An interesting finding for $Code_1$ (instances where the model fails to generate an output) is that there are no failed generations in $SET_{topp}$. All the generation failures occurred in $SET_{temp}$. Consequently, the coefficients for the two penalty values in this column are derived solely from the first set of tests, and no weighted average calculations were necessary. In addition, we can infer that the GPT model only fails to generate an output when the temperature is altered, indicating that temperature is the primary hyperparameter causing the model to malfunction. This observation aligns with the error messages mentioned in Section~\ref{sec:empiricalEvaluationSettings}, where the model consistently recommends lowering temperature in its error messages.

\begin{RQsTakeAwaybox}[RQ1 Results]
\textit{The variation in LLM hyperparameters does impact the code generated, with $temp$ leading to the maximum impact, followed by alternatively altering $top\_p$. $freq\_{pen}$ and $pres\_{pen}$ have a noticeable but weak impact on the code generation task.}
\end{RQsTakeAwaybox}



\subsection{RQ2 Analysis} \label{subsec:RQ2}
In order to identify the most positively impacting hyperparameter settings, we conduct a regression tree analysis~\cite{breiman2017classification}. A regression tree is built by recursively partitioning a dataset, such as the number of correct combinations generated, in a stepwise manner to create partitions that are as homogeneous as possible with respect to the target variable, $Code_0$ results, i.e., the correct results generated.

\begin{figure*}[htbp]
    \centering
    \includegraphics[width=0.85\textwidth]{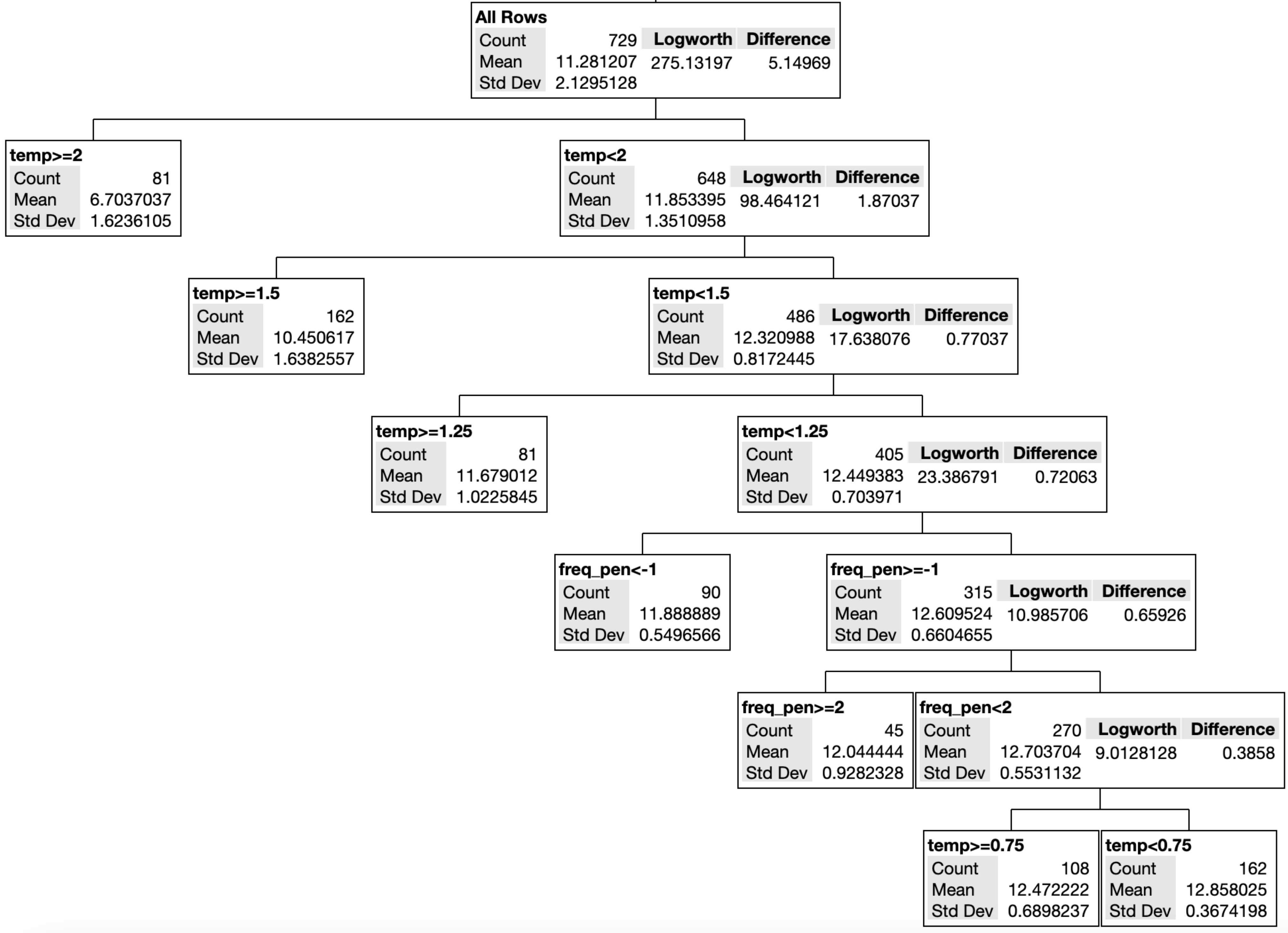}
    \caption{Regression Tree for $SET_{temp}$.}
    \label{fig:settemp_RegressionTree}
\end{figure*}

\begin{figure}[htbp]
    \centering
    \includegraphics[width=0.5\textwidth]{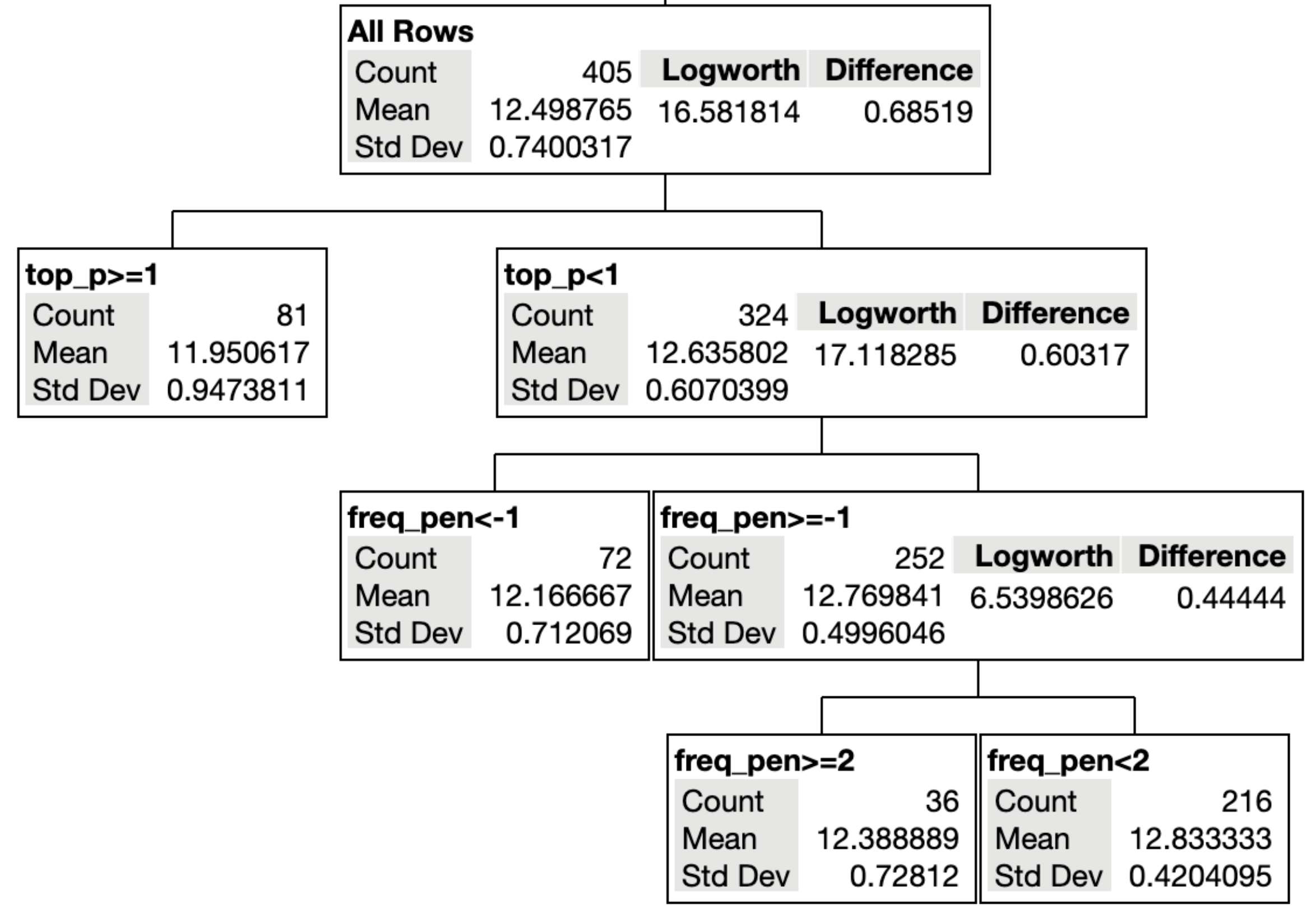}
    \caption{Regression Tree for $SET_{topp}$.}
    \label{fig:settopp_RegressionTree}
\end{figure}

Fig.~\ref{fig:settemp_RegressionTree} shows the regression tree for the 729 combinations in $SET_{temp}$, and Fig.~\ref{fig:settopp_RegressionTree} shows the regression tree for the 405 combinations in $SET_{topp}$.
At each level in the regression tree, one hyperparameter is selected, and the combinations are partitioned based on different values of that hyperparameter. The criterion for selection is to choose the hyperparameter that minimizes the standard deviation across the resulting branches. In other words, the hyperparameter that most significantly explains the variance in the number of correctly generated solutions out of 13 is selected. Each node in the tree displays the count (number of combinations), the mean and standard deviation of the total correctly generated solutions, and the difference between the smallest and largest sum observed in the partition. We stopped the partitioning after the difference was negligible (i.e., standard deviation less than 0.5). 

As shown in the regression tree of Fig.~\ref{fig:settemp_RegressionTree}, $temp$ is the most influential hyperparameter in $SET_{temp}$, followed by $freq\_{pen}$, with $pres\_{pen}$ leading to no noticeable impact on the correct solutions generated. Based on the regression tree, the best results are obtained by selecting the $temp<0.75$, i.e., [0, 0.5] values and $freq\_{pen} >= -1$ and $freq\_{pen} < 2$, i.e., [-1, 1.5] in our combinations, i.e., on an average 12.86 generated solutions out of 13 are correct for these combinations.  

As shown in the regression tree of 
Fig.~\ref{fig:settopp_RegressionTree}, $top\_p$ is the most influential hyperparameter in $SET_{topp}$. Similar to $SET_{temp}$ $freq\_{pen}$ has some influence, but $pres\_{pen}$ leads to no noticeable impact on the correct solutions generated. Based on the regression tree, the best results, i.e., on an average 12.83 generated solutions out of 13 are correct for $top\_p<1.0$, i.e., [0, 0.75] values and $freq\_{pen} >= -1$ and $freq\_{pen} < 2$, i.e., [-1, 1.5] in our combinations (which are exactly the same as Fig.~\ref{fig:settemp_RegressionTree}).

\begin{RQsTakeAwaybox}[RQ2 Results]
Based on our experiments, the combinations of $temp$ between 0.0 and 0.5 or $top\_p$ between 0.0 and 0.75, $freq\_{pen}$ between -1.0 and 1.5 and the default value of $pres\_{pen}$ (0.01) should lead to meaningful code generation results.     
\end{RQsTakeAwaybox}

%% file: Files/discussion.tex
\section{Discussion} \label{sec:discussion}

\begin{figure}[!htb]
    \centering
    \includegraphics[width=\columnwidth]{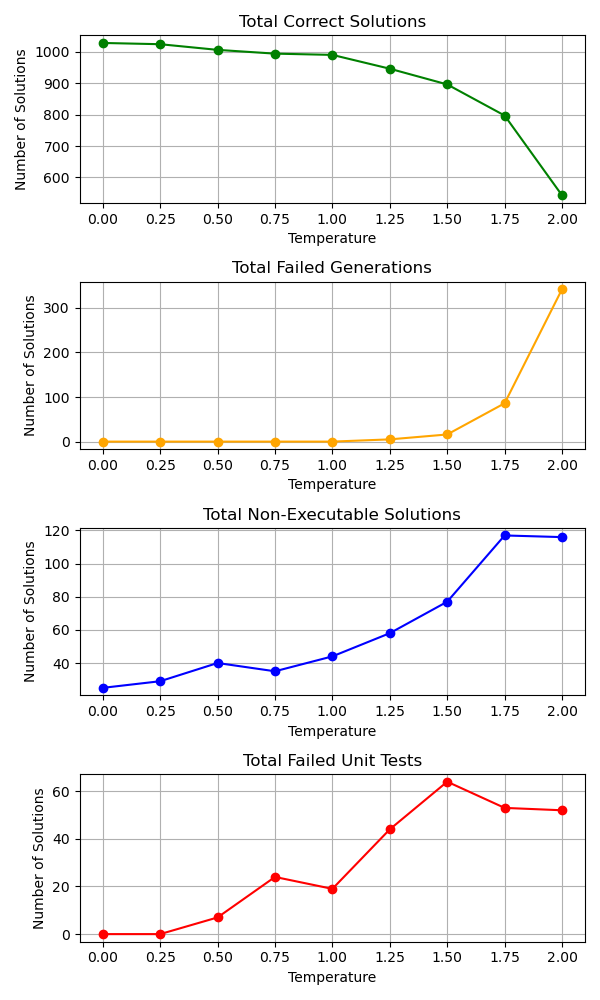}
    \caption{Line plot visualisation of the outcomes against temperature}
    \label{fig:temp_lines}
\end{figure}

\begin{figure}[!htb]
    \centering
    \includegraphics[width=\columnwidth]{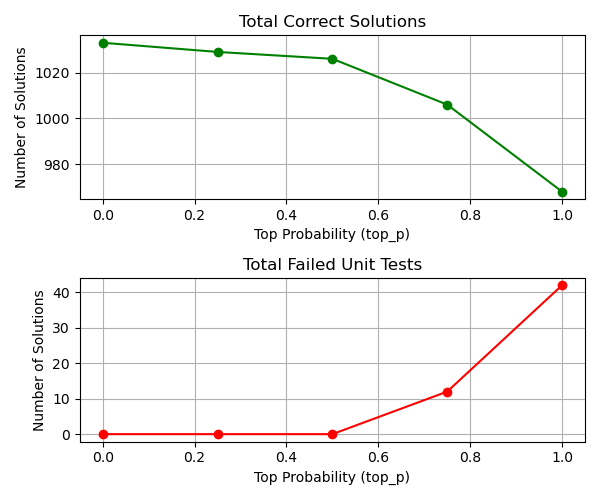}
    \caption{Line plot visualisation of the outcomes against top\_p}
    \label{fig:topp_lines}
\end{figure}

\begin{figure}[!htb]
    \centering
    \includegraphics[width=\columnwidth]{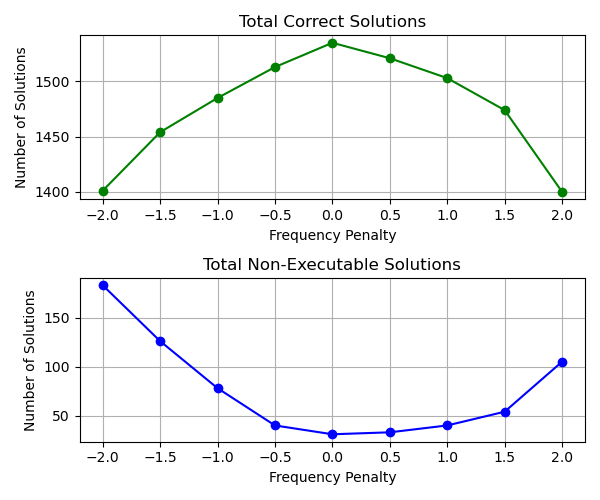}
    \caption{Line plot visualisation of the outcomes against frequency penalty}
    \label{fig:freqpen_lines}
\end{figure}

\begin{figure}[!htb]
    \centering
    \includegraphics[width=\columnwidth]{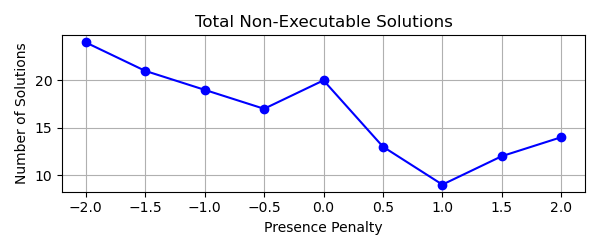}
    \caption{Line plot visualisation of the outcomes against presence penalty}
    \label{fig:prespen_lines}
\end{figure}

In this section, we delve deeper into the discussion of the results and report on additional analyses. Using the results from RQ2, we wanted to further explore how the hyperparameters impact the respective outcomes. We focused only on the relationships that were picked up by the individual regression trees (not all regression trees are visualised in the paper in detail due to space constraints). Figs.~\ref{fig:temp_lines}, \ref{fig:topp_lines}, \ref{fig:freqpen_lines} and \ref{fig:prespen_lines} show the line graphs for different outcomes of varying different hyperparameters.
For example, we only visualised the effect of the $pres\_pen$ for outcome $Code_2$ for $SET_{topp}$, as the other regression trees did not split using this hyperparameter for any other outcomes. In addition, as $freq\_pen$ mainly impacted $Code_0$ and $Code_2$ for both sets, we decided to visualise this hyperparameter by considering both sets in conjunction. In order to produce these line graphs, we aggregated the total number of solutions generated for each outcome for each hyperparameter. For $temp$ and $top\_p$, we summed the total solutions for their respective sets of data. For $freq\_pen$, we aggregated the total solutions from both sets of data (as it was present in the regression trees from both sets). Finally, for $pres\_pen$, we aggregated the total solutions from just $SET_{topp}$ and not $SET_{temp}$. We also summarise these results for convenience in Table~\ref{tab:hyperimpact}.

\begin{table}[!htb]
\caption{Impact of Each Hyperparameter on Each Outcome Type}
\label{tab:hyperimpact}
\centering
\begin{tabular}{|c|c|c|c|c|}
\hline
Outcome Types      & 0 & 1 & 2 & 3 \\ \hline
Temperature        & \textbf{high} & \textbf{high} & \textbf{high} & \textbf{high}  \\ \hline
Top Probability    & \textbf{high} & none  & low  & \textbf{high}  \\ \hline
Frequency Penalty  & moderate & low  & \textbf{high} & low  \\ \hline
Presence Penalty   & low  & low  & moderate & low  \\ \hline
\end{tabular}
\end{table}


In Fig.~\ref{fig:temp_lines}, we can see how each outcome changes as $temp$ increases. For outcome $Code_0$ (instances when the solution is correct), the curve remains fairly constant until a temperature of 1.0 and then declines afterwards. The results are consistent with our regression trees in Fig.~\ref{fig:settemp_RegressionTree}, as we recommend maintaining the $temp$ between 0.0 and 0.5. These results are consistent with previous studies on hyperparameters as well~\cite{ouyang2023llm, bhavya2022analogy}, which show that LLMs perform better when configured to be more deterministic. We further note that $temp$ values beyond 1.50 yield significantly worse results, and hence, such high $temp$ values, while maybe suited for more creative tasks, are not well suited for code generation.

Similarly, $top\_p$ in Fig.~\ref{fig:topp_lines} has a similar pattern to temperature for both outcomes $Code_0$ and $Code_3$. As $top\_p$ is increased, the number of correct solutions ($Code_0$) gradually declines. The number of failed unit tests ($Code_3$) is constant until 0.5 and then starts to increase linearly. Hence, it is recommended to maintain $top\_p$ values low. We also note that the default value of $top\_p$ maintained by OpenAI API is 1.0, which is higher than our recommendations. Hence, this shows the value of our research, where hyperparameter tuning for specific SE tasks can yield better results than relying on default values for generic text generation tasks. For $top\_p$, very high values close to 1.0 should be avoided for code generation.

For $freq\_pen$ in Fig.~\ref{fig:freqpen_lines}, the graph for $Code_0$ shows a peak at 0.0, indicating that this penalty yields the highest number of correct solutions, with a gradual decline as the penalty moves away from 0 in either direction. This aligns with our findings from regression analysis, as values between -1 and +1.5 demonstrated better performance. The second graph ($Code_2$) shows a U-shaped curve, with the fewest non-executable solutions occurring near a frequency penalty of 0 and an increase in non-executable solutions as the penalty deviates further from 0. Hence, the values too low, e.g., less than -1.0 and too high, e.g., higher than 1.5 should be avoided for code generation.

Lastly, the line graph shows a more erratic pattern for $pres\_pen$ in Fig.~\ref{fig:prespen_lines}. However, as we discussed in RQ2 results, $pres\_pen$ is not a very impactful hyperparameter. The general trend remains as expected, where the number of failed solutions ($Code_2$) decreases as the penalty value increases. Hence, $pres\_pen$ to be avoided could be the ones close to -2.0.

\begin{figure}[!htb]
    \centering
    \includegraphics[width=\columnwidth]{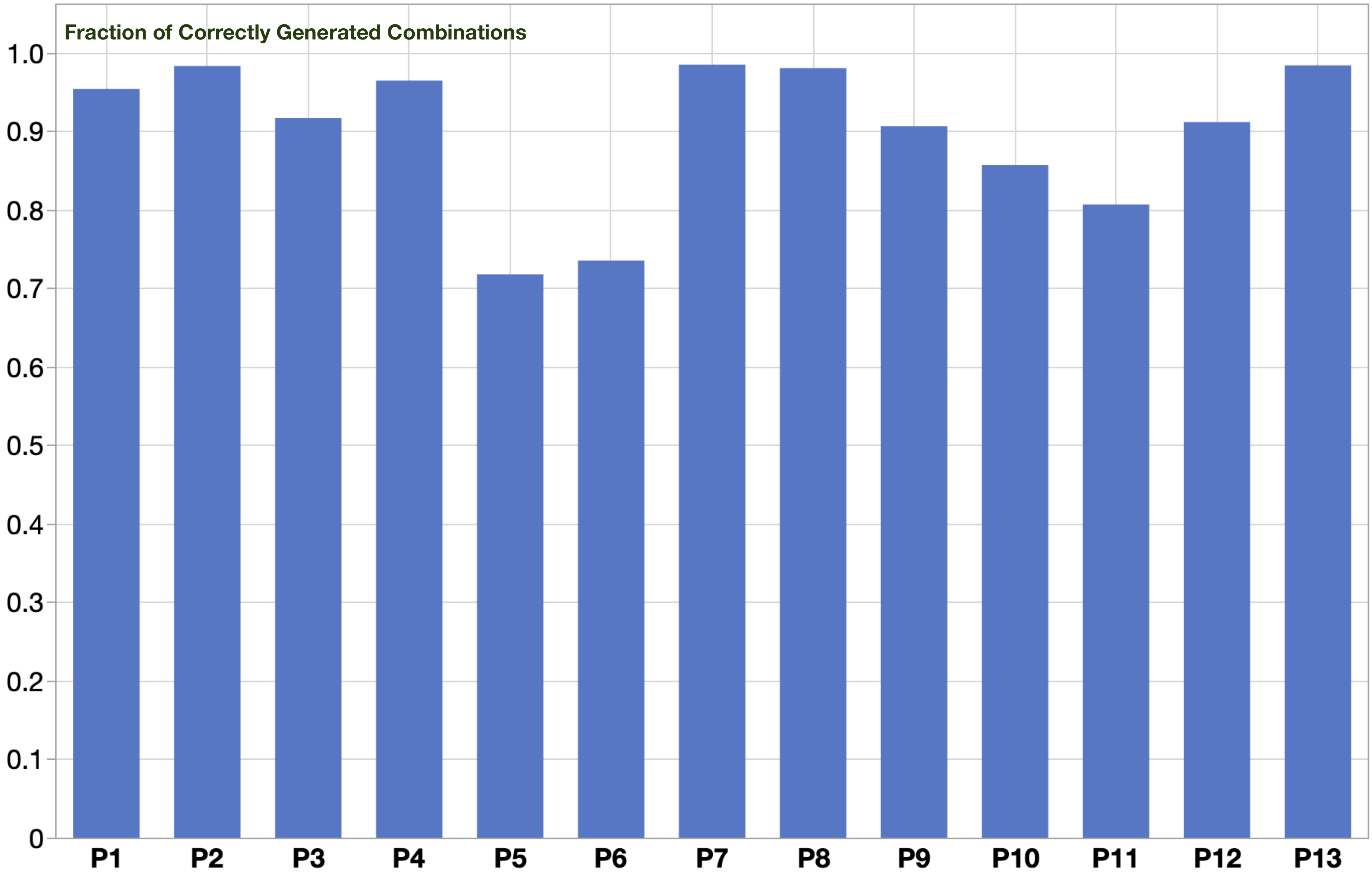}
    \caption{Fraction of Task Combinations Correctly Generated}
    \label{fig:TasksCorrectness}
\end{figure}

We further analysed the results for different tasks from the 13 tasks analysed in this study. Our results (Fig.~\ref{fig:TasksCorrectness}) show that overall, the average correct solutions generated are relatively consistent and high across most tasks (more than 90\% correct solutions), except P5 (71.9\% correct solutions), P6 (73.6\% correct solutions), P10 (85.8\% correct solutions), and P11 (80.8\% correct solutions), i.e., these four tasks contribute to the most incorrect solutions ($Code_1$, $Code_2$ and $Code_3$) in Fig.~\ref{fig:bargraph}. The tasks that did not work well share common factors of increased complexity, involving multiple steps, specific libraries, advanced mathematical operations, and recursion (see description in Table~\ref{tbl:Programs}). P5 involves reading from a text file, summing groups of integers, and identifying the group with the largest sum. This task has multiple steps: file handling, parsing data, summing groups, and comparing sums. In P6, although the FizzBuzz problem is relatively straightforward, it involves conditional logic with multiple branches, which might be tricky for the model if not handled properly. P10 requires using NumPy, a specific library, and performing complex mathematical operations (determinant and eigenvalues), which adds a layer of complexity. In P11, writing a recursive function to generate the Fibonacci sequence involves understanding recursion, which can be challenging for code generation models if not properly framed. This also shows that while LLMs consistently perform well for code generation of relatively straightforward and simple problems, they might falter in generating programs that require a higher level of complexity~\cite{jiang2024survey}. While the model did generate correct solutions for these problems in more than 70\% cases, the impact of hyperparameters, prompts and potentially step-by-step guidance to LLMs should be considered when generating code for relatively complex problems.

%% file: Files/threats.tex
\section{Threats to Validity} \label{sec:threats}
\sectopic{Internal Validity.} The main threat to internal validity is related to the hyperparameter increments. The selection of hyperparameter increments (e.g., 0.25 for $temp$ and $top\_p$) might not capture the optimal settings accurately. Different increments might potentially reveal more nuanced effects. To mitigate this issue, we conducted additional experiments at intermediate values, which were not systematically included in the main study. These supplementary tests did not indicate significant changes in the outcomes, suggesting that the chosen increments were reasonable. Furthermore, as demonstrated in the results for RQ2, the optimal ranges are sufficiently broad and align well with the selected increments, thereby supporting the validity of our approach.

\sectopic{Construct Validity.}
The metrics used to evaluate the correctness and functionality of the generated code (passing the unit tests generated by the research team) might not capture all aspects of code quality, such as readability, maintainability, or efficiency~\cite{NDUKWE2023111524}. While we tried to generate unit tests (using assertions) for basic input-output pairs, more tests could be required in future studies.

\sectopic{External Validity.} The study focuses on a specific model (GPT-3.5), a specific programming language (Python), and specific types of programming tasks (13 tasks). The results may not generalise to other LLMs or future versions of the model, other programming languages and other types of programming tasks that might have different sensitivities to hyperparameter settings. While we tried to select programming tasks of different types, more extensive studies are nonetheless required with more LLMs, more programming languages and a larger task base.

%% file: Files/related.tex
\section{Related Work}
\label{sec:RelatedWork}

Since OpenAI first launched ChatGPT in 2022\footnote{https://hbr.org/2022/12/chatgpt-is-a-tipping-point-for-ai}, it has garnered significant public attention, not only for its role within the GPT series but also for the advancements in LLMs overall as well. However, code-related research on LLMs has long been a focus for software engineering researchers~\cite{yang2024if,daye2024using}.
Some research papers have shown the exceptional performance of LLMs, including works focused on code generation~\cite{poesia2022synchromesh,bareiss2022code}, code review~\cite{lu2023llamareviewer,zhang2024detecting}, debugging~\cite{lee2024unified,jiang2024training}, and code translation~\cite{pan2023understanding,Pan_2024}.

Meanwhile, several limitations of LLMs have been identified in existing analysis. Solving complex coding problems remains a significant challenge for these models. Yan et al.\cite{10298505} demonstrate this by showing a low code pass rate of only 25\% for ChatGPT on complex problems. Additionally, Sakib et al.\cite{sakib2023extending} highlight the model’s poor performance in improving solutions and debugging based on human feedback, and Sun et al.~\cite{sun2023automatic} reveal limitations in code summarization of LLMs. These findings collectively indicate that the solution of comprehensive code understanding and generation still remains out of reach for current LLMs.

In order to address these challenges, various research efforts have been made to enhance the effectiveness of LLMs in code generation. One approach is to divide complex problems into simpler tasks, incorporating self-planning within the code generation framework to enhance problem-solving capabilities~\cite{zhang2023planning,jiang2023selfplanning}. Li et al.\cite{li2023think} and Mu et al.\cite{mu2023clarifygpt} have introduced new frameworks to refine ambiguous requirements and provide selective solutions for complex problems. Fine-tuning LLMs on private datasets is another solution that has been widely considered in previous studies~\cite{zan2023privatelibraryoriented, chen2021evaluating}. Beyond improving the coding abilities, some studies have focused on enhancing other aspects of software development to achieve a more holistic improvement of LLMs. Du et al.\cite{du2023resolving} proposed an interactive methodology to improve the resolution of crash bugs. Chen et al.\cite{chen2023teaching} developed techniques for teaching LLMs to self-debug their solutions without human feedback. Lastly, Pan et al.~\cite{pan2023stelocoder} aimed to improve code translation by introducing an LLM specifically designed to translate various programming languages into Python.

Hyperparameter initialization is a crucial step in configuring LLMs. However, there has been limited research on the impact of different hyperparameter configurations. Zhang et al.~\cite{zhang2023using} revealed that hyperparameter settings significantly influence a model’s behaviour and performance. Hyperparameters, which can be adjusted before training or during training, fine-tune the model’s behaviour and control the generation process. Some hyperparameters such as the learning rate and top probability play a significant role in determining the final performance of LLMs. Ouyang et al.~\cite{ouyang2023llm} explored the effects of the temperature hyperparameter on the non-determinism of ChatGPT in code generation. Their results, which only report temperature values of 0, 1, and 2, suggest the need to investigate how smaller increments in temperature might affect correctness instead of only using extreme values. In terms of prompt-related hyperparameters, Bhavya et al.~\cite{bhavya2022analogy} conducted limited experiments with frequency penalty and presence penalty on deterministic and non-deterministic behaviours. These studies drive the need for a more comprehensive analysis of prompt-related hyperparameters to fully understand their impact on LLM performance, which is the focus of this study.

%% file: Files/conclusion.tex
\section{Conclusion} \label{sec:conclusion}
Our study focused on the systematic investigation of hyperparameter tuning of the GPT 3.5 model via API for Python code generation. In our study, we performed systematic tuning of four key hyperparameters of LLMs, namely \textit{temperature}, \textit{top probability (top\_p}), \textit{frequency penalty} and \textit{presence penalty}. Using step-wise increments in the values of these four hyperparameters, we generated 1,134 combinations of these hyperparameters and generated Python code for 13 tasks using each of these 1,134 combinations, i.e., generated and compared 14,742 Python functions. We used assertions and the \textit{unittest} library in Python to validate the functionality of the generated code. We further classified the failed cases into three categories: (i) fails to generate a solution ($Code_1$); (ii) the generated solution is not executable ($Code_2$); and (iii) the generated solution fails (one+) unit tests ($Code_3$). 

In our evaluation, we provide empirical evidence to the anecdotal assertions that hyperparameters do impact LLMs' code generation performance and that certain hyperparameters influence code generation more than others. We also identified optimal and subpar hyperparameter values for code generation in Python. Temperature has the highest impact on all outcomes. Setting this hyperparameter to low (below 0.5) showed improvements in all aspects. Top probability (top\_p) moderately influences failed unit tests, so setting that to lower values (below 0.75) has improved correctness. Frequency penalty increases the likelihood of non-executable code being outputted by the GPT model. Setting this hyperparameter between -1 and +1.5 leads to better results. Finally, the presence penalty only marginally impacts code generation, leading to non-executable code in some cases. Setting it above -1 will show slight improvements. Some of these settings are different from the default settings proposed by OpenAI API for generic text generation. Hence, our results are valuable, as the default settings might not lead to optimal results. 

In the future, we aim to expand the study to include a wider evaluation of more LLMs, programming languages, and programming tasks. We also plan to test the hyperparameter settings in real-world software development environments to observe their practical impacts and gather feedback from developers. We further plan to expand the scope of our study beyond code generation to other SE automation tasks via LLMs, e.g., requirements analyses, test case generation, and model generation.